\documentclass[aps,prl,twocolumn,floatfix]{revtex4}
\usepackage{graphicx}
\usepackage{mathptmx}      
\usepackage{graphicx,psfrag}
\usepackage{amsmath}
\begin{document}

\title{Density-dependence of functional development in spiking cortical networks grown {\it in vitro}} 

\author{Michael I. Ham}
\affiliation{
Center for Nonlinear Studies and Applied Modern Physics P-21, Los Alamos National Laboratory, Los Alamos NM 87544 Tel.: +1-505-565-6230 mikeh@lanl.gov}
\author{Vadas Gintautas}
\affiliation{
              Center for Nonlinear Studies and Applied Mathematics and Plasma Physics, Los Alamos National Laboratory, Los Alamos NM 87544}
\author{Marko A. Rodriguez}
\affiliation{
              Center for Nonlinear Studies and Applied Mathematics and Plasma Physics, Los Alamos National Laboratory, Los Alamos NM 87544}
\author{Ryan A. Bennett}
\affiliation{
              Center for Network Neuroscience and Department of Physics,
              University of North Texas, Denton TX 76203}
\author{Cara L. Santa Maria}
\affiliation{
              Center for Network Neuroscience and Department of Biology,
              University of North Texas, Denton TX 76203}
\author{Lu\`is M.A. Bettencourt}
\affiliation{
              Theoretical Division,
              Los Alamos National Laboratory, Los Alamos NM 87544 and
              Santa Fe Institute, 
              1399 Hyde Park Road 
              Santa Fe NM 87501}

\begin{abstract}
During development, the mammalian brain differentiates into specialized regions with distinct functional abilities.  
While many factors contribute to functional specialization, we explore the effect of neuronal density on the development of neuronal interactions {\it in vitro}. 
Two types of cortical networks, dense and sparse, with $50,000$ and $12,000$ total cells respectively, are studied.   
Activation graphs that represent pairwise neuronal interactions are constructed using a competitive first response model. 
These graphs reveal that, during development {\it in vitro}, dense networks form activation connections earlier than sparse networks.
Link entropy analysis of dense network activation graphs suggests that the majority of connections between electrodes are reciprocal in nature.
Information theoretic measures reveal that early functional information interactions (among 3 cells) are synergetic in both dense and sparse networks.  
However, during later stages of development, previously synergetic relationships become primarily redundant in dense, but not in sparse networks. 
Large link entropy values in the activation graph are related to the domination of redundant ensembles in late stages of development in dense networks. 
Results demonstrate differences between dense and sparse networks in terms of informational groups, pairwise relationships, and activation graphs.  
These differences suggest that variations in cell density may result in different functional specialization of nervous system tissue {\it in vivo}. 
\end{abstract}

\maketitle

\section{Introduction}
\label{intro}
The mammalian brain is a remarkable structure composed of many specialized regions and types of cells. 
Despite organizational differences in neural tissue, the basic functional units of the nervous system, neurons, are generally similar across tissues, as are methods of forming and modifying synaptic connections between them (e.g. spike-timing dependent plasticity \cite{markram_1997}, \cite{bi_1998}).
 Thus, functional specialization of brain regions is a function of neuron specialization, (e.g. excitatory or inhibitory), ratios of neurons to neuroglia, synaptic density, learning, and many other factors that are not yet well understood.

 Though the functional role of neurons in different types of tissues can be similar, neuronal density can vary dramatically.  For example, neuronal density in human fascia dentata is $\approx3.2 X 10^5$ neurons/mm$^3$ \cite{kim_1990} while cortical tissue density is $\approx3.4 X 10^4$ neurons/mm$^3$ \cite{anderson_1996}, nearly an order of magnitude difference. 

 A previous study of network development {\it in vitro} demonstrated that network bursting (when a large majority of  neurons fire in a coordinated pattern) and spiking patterns are affected by neuronal density \cite{wagenaar_2006}. Since spike and burst activity are observable results of neuronal interaction (functional connectivity) they can be used to infer neuronal relationships \cite{pfister_2006}, \cite{bcourt_2007}, \cite{tang_2008}, \cite{bettencourt_2008}. 

In this work the effect of density on functional units is explored using dissociated cortical tissue developing {\it in vitro} on microelectrode arrays.  
These are well established models of neuronal interaction \cite{paninski_2003}, \cite{beggs_2004}, \cite{feinerman_2007}, though obvious constraints and limitations must be considered when attempting to extrapolate between {\it in vitro} and {\it in vivo} structures \cite{corner_2002}, \cite{marom_2002}.
To gain new insight into functional connectivity in developing networks, we analyze the coordinated electrophysiological activity of groups of two and three spike trains, each representing an integration of all of the action potentials recorded at a single electrode. Each electrode may capture the activity of a single neuron, or less frequently, incorporate signals from several cells. 

Activation between pairs of electrodes is inferred using a competitive first response model, whereby directional links are derived from spike train data. 
These pairwise links estimate the probability that activity at one electrode causes activity at another. 
The set of all such links forms a network activation graph where vertices represent recording electrodes and weighted edges represent a dependent activation probability. 
We use entropy based link analysis (link entropy) to characterize connectivity between electrodes. 
Information theoretic measures applied to ensembles of $3$ electrodes reveal functional information structures. 
Such structures are characterized using an information overlap method that reveals whether interactions are synergetic (more information is obtained from the measurement of two electrodes together conditioned on the third compared to measuring them separately), redundant (less information is obtained from the measurement of two electrodes together conditioned on the third compared to measuring them separately) or independent. 
Results show that functional structures in networks are strongly influenced by neuronal density and suggest that varying cell density is a potential strategy for differentiating tissue functionality.

\section{Methods}
\label{Methods}
We analyze developmental activity patterns in eight cultured cortical networks growing on microelectrode arrays. Networks were recorded \cite{wagenaar_2006} on days ranging from $3$ to $36$ days {\it in vitro} (DIV, days after plating). 
Half the networks studied are sparsely seeded ($625$ cells/ml) the other half densely seeded ($2,500$ cells/ml).
 
\begin{figure*}[htbp]
    \includegraphics[width=.95\textwidth]{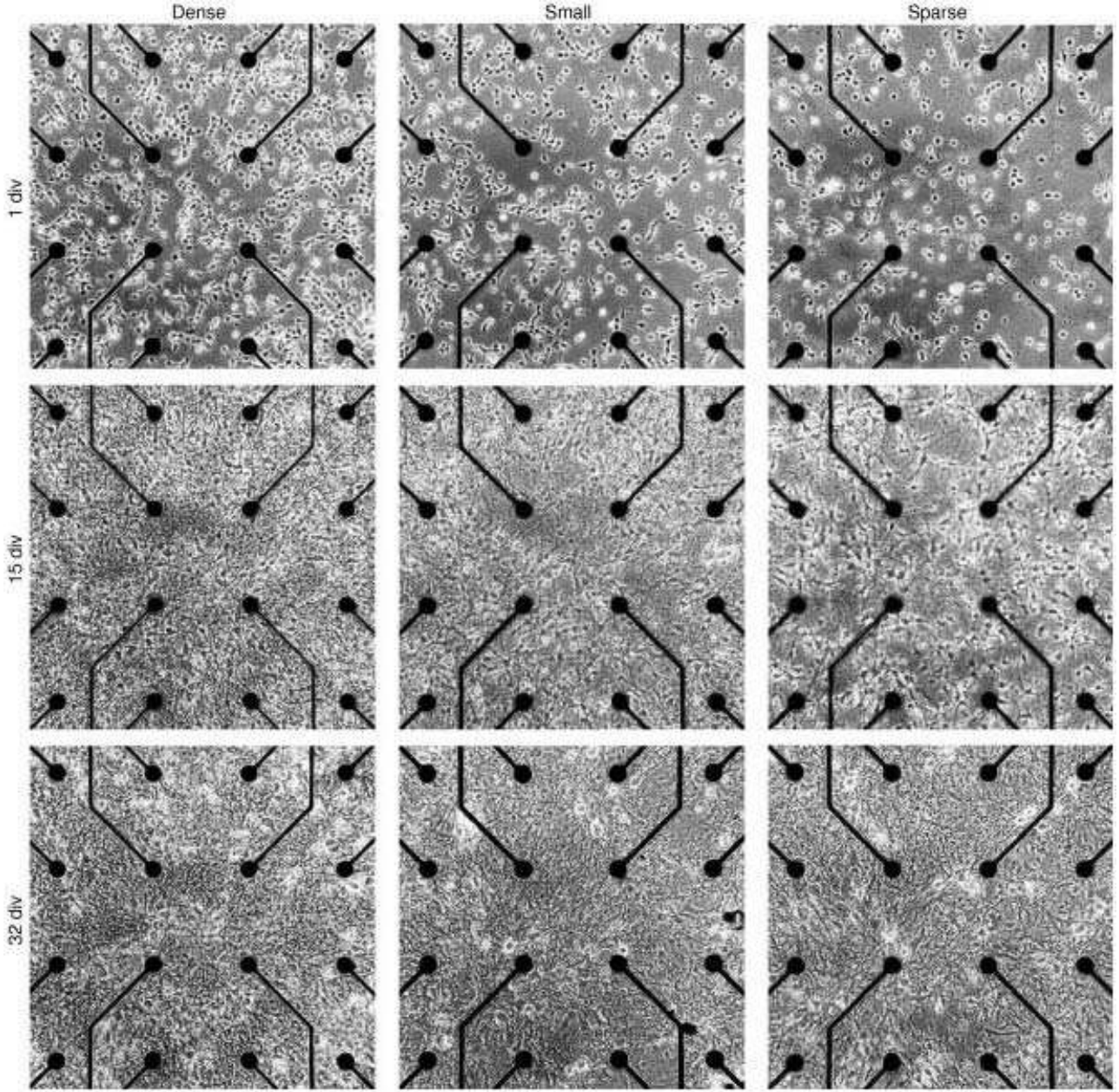}
    \caption{Neurons and neuroglia growing on microelectrode arrays. Dense (left column) and sparse (right column) networks are depicted at $1$, $15$ and $32$ days {\it in vitro} (DIV). Dark circles are electrodes and dark lines are integrated wires that transmit action potential activity from electrodes to amplifiers (not pictured).  Figure was previously published by Wagenaar et al. \cite{wagenaar_2006} and is used with permission from the authors under the license of the publication \label{fig:f1}}
\end{figure*}

Cells (neurons and support glia) are removed from embryonic mice and their existing tissue structure is dissociated mechanically and enzymatically. 
After cells are seeded on a microelectrode array, they form new connections and self-organize into spontaneously active networks \cite{grossbook_1994}, \cite{maeda_1995}. 
Figure \ref{fig:f1} depicts a representative dense and sparse network at $1$, $15$ and $32$ DIV. 
Experiments and data collection were performed by Wagenaar et al. \cite{wagenaar_2006} and made publicly available for analysis. 
Detailed information about culturing, plating, and feeding techniques are provided in this reference.

\subsection{First response model}
\label{frmodel}
We build a model of network connectivity using spike timing correlations to study the evolution of pairwise neural interactions during development. 
The model is based on the assumption that, within a small network, all activity preceding a spike within a biologically plausible time window contributes to its firing.  
Any given spike is assumed to be correlated with the ignition of the next spike to fire within $1-10$ ms. 
This temporal window is selected such that the first spike has enough time to influence the production of the second, but not so long that its effect will have faded \cite{nakanishi_1998}, \cite{bi_1998}. 
All spikes collected at a single electrode form a single spike train.
Therefore, the number of spike trains is equal to the number of active channels (Table \ref{table:t1}) and functional connections discussed here are made between electrodes. 

\begin{table}
\begin{tabular}{l c l}
Network & Electrodes & Recording days \\
\hline
D1 & 56 & 4-7,9-26,28,31-35\\
D2 & 56 & 4-26,28,31-35\\
D3 & 56 & 4-26,28,31-35\\
D4 & 56 & 4-26,28,31-35\\
S1 & 58 & 4-7,10,11,13,14,17,19,21,22,24-26,28,31,33,34  \\
S2 & 58 & 4-7,10,11,13,14,17,19,21,22,24-26,28,31,33,34  \\
S3 & 58 & 4-7,10,11,13,14,17,19,21,22,24-26,28,31,33,34 \\
S4 & 58 & 4-8,10,12-14,17,18,20,21,31-35 \\
\end{tabular}
\caption{Network: D1-D4 are densely seeded and S1-S4 are sparsely seeded. Electrodes: number of electrodes with action potential activity. Recording days: days {\it in vitro} a network was recorded. \label{table:t1}}
\end{table}

Recent analysis of similar recordings in which individual neurons at an electrode were discriminated (spike sorted) shows that, for plating densities comparable to those in the data presented here, typically no more than 2 neurons are observed at any given electrode \cite{ham_2008}.
 
\begin{figure}[htbp]
    \includegraphics[width=.45\textwidth]{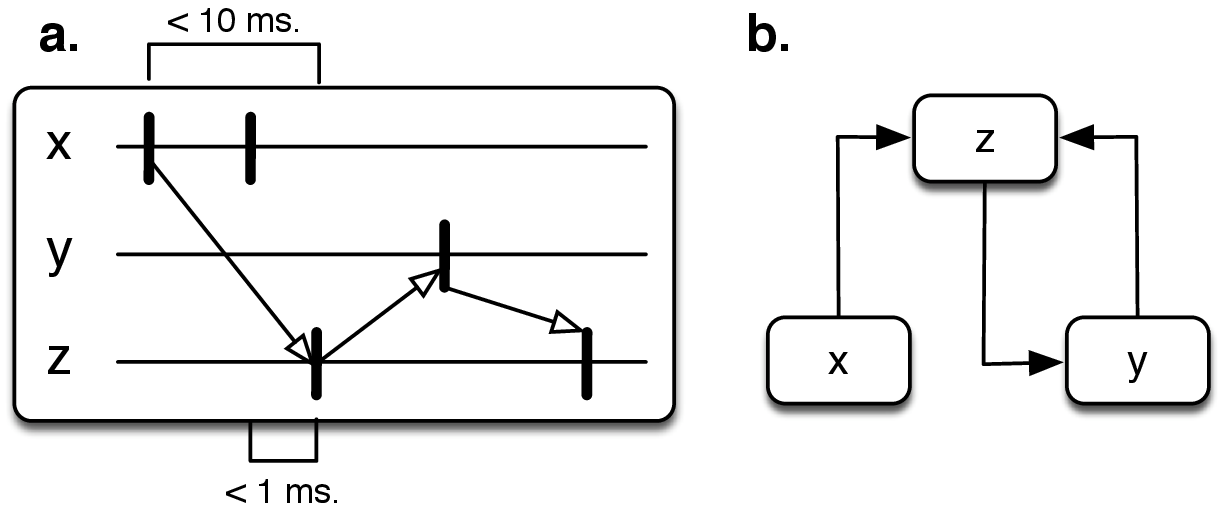}
     \caption{a) Activation links are made when activity at an electrode is the first to occur within a $1-10$ ms window after another electrode.
Note that no link is made from the second spike in $X$ since the time window between it and $Z$ is less than $1$ ms and it and $Y$ is greater than $10$ ms.    
b.) Pairwise activation graph constructed from data in part a.   
Edges are directed and weights increase as connections are observed. 
Edges are normalized by the lesser of the total spikes observed at either of the connected electrodes.   
Activation networks represent activation flow and are not representative of the physical connections between two electrodes \label{fig:f2}}
\end{figure}

The first response model provides a method for estimating whether a spike produced at one electrode initiates a spike at another. 
Each time a spike at an electrode is the first to fire within $1-10$ ms after a spike at another electrode (Fig. \ref{fig:f2}a), the link (edge) from the first to the second is incrementally increased. This results in a weighted and directed edge between electrodes (Fig. \ref{fig:f2}b).

Edge weights are calculated for all electrode pairs. 
Edges represent inferred activation links in the network and are not necessarily representative of actual synaptic connection between neurons recorded at any two electrodes. 
Edges are denoted by $x_{ij}$ , where $i$ is the first electrode to fire and $j$ the second.
When pairwise edges are studied individually, they are normalized by the lesser of the total spikes recorded at $i$ or $j$, giving an estimate of the probability that $i$ leads $j$. Normalized frequency edges $x_{ij}$ and $x_{ji}$  are denoted $X_n$ and $Y_n$ respectively.

\subsection{Activation graph}
\label{acgraph}
The combination of all weighted directed edges between electrodes forms an activation graph, which is a weighted directed graph, with each edge normalized between zero and unity. 
This graph estimates the activation probability between all electrodes. 
New activation graphs are created for each network on each recording day.

\subsection{Information ensembles of order 2}
\label{io2}

Neuronal activation along pathways in the brain is generally a sequential process. 
Activity in one area can ignite activity in another on large scales (tissue structure, e.g. \cite{spitsyna_2006}) as well as small scales (neuronal networks, e.g. \cite{ham_2008}). 
To characterize the influence a target electrode has on all other electrodes, we apply a link entropy analysis of the activation graphs. 
Link entropy is calculated for each electrode and measures the uncertainty of which electrodes will be activated by the target electrode.  
For example, in Fig. \ref{fig:f2}b, electrode $X$ activates only $Z$. 
Since there is no uncertainty about which electrode $X$ activates, the link entropy is $0$. 
If $X$ activated more than one other electrode, uncertainty in activation and link entropy increase.
In a fully connected network with $N$ electrodes, there are $N-1$ potential pathways (edges) originating from each electrode. 
Edges ($x_{ij}$) originating from a target electrode ($i$) in the weighted directed activation graph are used to estimate the probability $p(x_{ij})$:

\begin{figure}[htbp]
    \includegraphics[width=.45\textwidth]{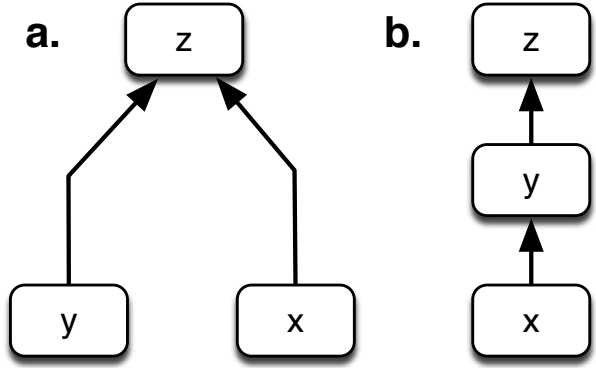}
     \caption{
a) Synergetic relationship where $X$ and $Y$ together provide more information about the state of $Z$ than they do individually, as in the case of a logic gate. 
b) Redundant relationship, where $X$ and $Y$ together provide less information about the state of $Z$ than they do individually, as in the case of a Markov chain   
     \label{fig:f3}}
\end{figure}

\begin{equation}
    p(x_{ij})=\frac{x_{ij}}{\sum_{j=1}^{N-1} x_{ij}}
\end{equation}
 
The link entropy of the target electrode ($i$) in an activation graph with N active electrodes is:

\begin{equation}
    H_i = -\sum_{j=1}^{N-1} p(x_{ij}) log_2(p(x_{ij}))
\end{equation}
 
This measure represents the uncertainty (in bits) of the potential activation paths originating from a target electrode. 
Link entropy values range from $0$ (no uncertainty; activation flows along only one path with probability of $100\%$) to a maximal value of $log_2(N-1)$. The latter occurs only when (non-zero) probabilistic connections between the target and all other $N-1$ electrodes in the network are equal.

\subsection{Informational ensembles of order $3$}
\label{order 3}

\begin{figure*}[htbp]
    \includegraphics[width=.95\textwidth]{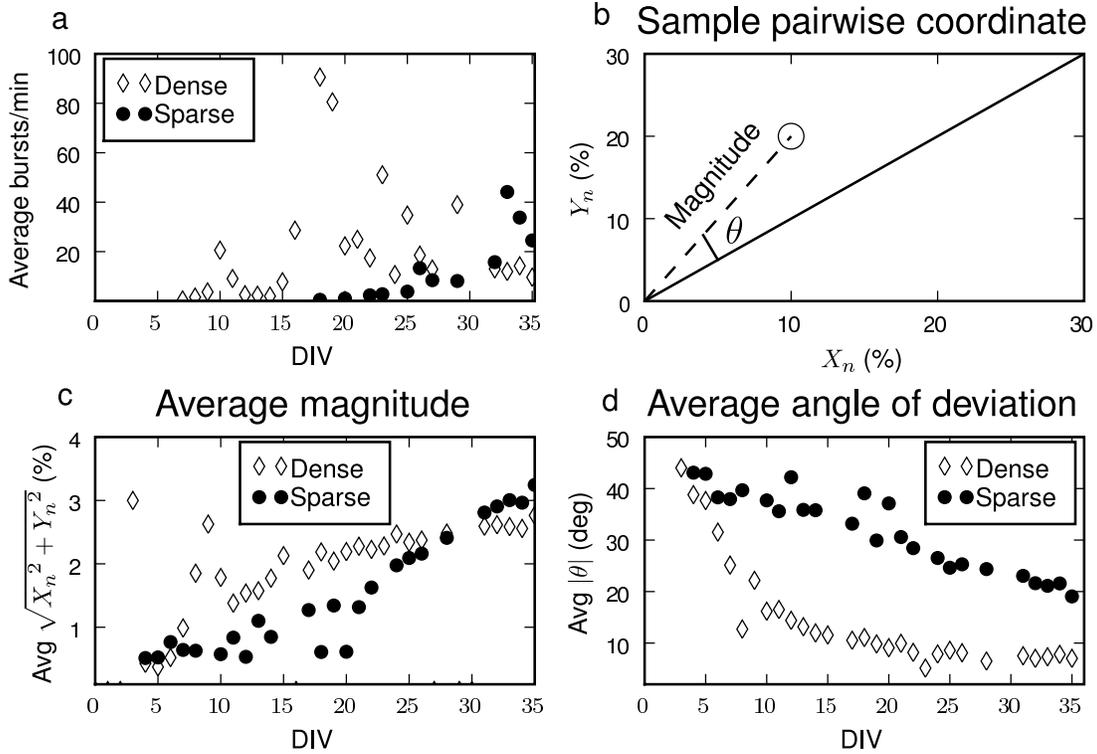}
     \caption{Comparison of network bursts and pairwise activity in dense and sparse networks. Four dense (diamonds) and four sparse (circles) are averaged together. a) Average number of network bursts per minute per day for dense and sparse networks.  Multiple network bursts per minute occur around $9$ DIV in dense networks and $22$ DIV in sparse networks. Days with no network bursts are not plotted. b) Example of a normalized pairwise interaction. 
$X_n$ is the probability that neuron $Y$ is the first to respond to neuron $X$ and vice-versa for $Y_n$. 
For all coordinates not equal to [0,0], the coordinate magnitude and angle of deviation from the diagonal $X_n$ = $Y_n$ (solid line) is computed. 
c) Time evolution of the average magnitude across all study networks in dense and sparse networks. Note that average magnitudes are relatively small due to the fact that very few pairs lead one another a majority of the time.  
Magnitude in dense networks increases faster, but sparse networks exhibit higher vales after 30 DIV. 
d) Average angle (degrees) of deviation in dense and sparse networks. 
Throughout maturation, sparse networks show greater deviation from the diagonal than dense ones. 
In both cases deviation decreases with age, but deviation in dense networks decreases fasters \label{fig:f4}}
\end{figure*}

Information theoretic measures applied to ensembles of $3$ electrodes are used to estimate functional relationships, beyond simple directed connectivity, in developing networks. 
At this point, probabilities are no longer based on activation graph links between electrodes. Rather, each electrode can assume two states -- spiking or not spiking. 
In the convention established in Reike et al. \cite{reike_1997}, see also \cite{bcourt_2007}, information theoretic quantities are measured only during network bursts, where the vast majority of all coordinated interactions occur. More details about network burst detection are given in \cite{ham_2008}. The average number of detected network bursts per minute across all dense and sparse networks on all recorded DIV is shown in Fig. \ref{fig:f4}a. Bursts are digitized by dividing time into $10$ ms bins. If activity occurs during a bin, the bin is assigned a $1$.  Otherwise the bin is assigned a $0$. Probabilities are computed by normalizing the number of spiking and not spiking bins by the total number of bins. Joint probabilities can also be computed in the same way for multiple electrodes.

After digitization, all groups of spike train triplets are examined for redundant or synergetic relationships per the convention established in \cite{bcourt_2007}, \cite{schneidman_2003}.
Specifically, for three electrodes ${X,Y,Z}$, we consider the conditional mutual information between $X$ and the state of two other electrodes $Y$ and $Z$ (\cite{cover_1991}):

\begin{equation}
    I(X;\{Y,Z\})=\sum_{x,y,z} p(x,y,z) log_2 \frac{p(x,y,z)}{p(x)p(y,z)}.
\end{equation}
 
Subtracting this quantity from the sum of $I(X;Y)$ and $I(X;Z)$ reveals the informational nature of the ensemble \cite{cover_1991}. 
For example, in a synergetic ensemble (see Fig. \ref{fig:f3}a), more information is gained when considering $Y$ and $Z$ together rather than separately: $I(X;{Y,Z}) > I(X;Y) + I(X;Z)$.
However, in a redundant ensemble (see Fig. \ref{fig:f3}b), less information is gained when considering $Y$ and $Z$ together rather than separately: $I(X;{Y,Z}) < I(X;Y) + I(X;Z)$.  If $Y$ and $Z$ are independent, $I(X;{Y,Z}) = I(X;Y) + I(X;Z)$.
Therefore, we can use the measure $R(X,Y,Z)=I(X;Y) + I(X;Z)-I(X;{Y,Z})$ to characterize the functional informational nature of an ensemble.  $R$ can be positive (redundant), negative (synergetic) or zero (independent).

To test for statistical significance, experimentally obtained $R$ values were compared to values obtained from a Poisson null model with the same spiking rates as in the experiment \cite{bcourt_2007}. In this model, experimentally obtained spike trains are used to form a randomly distributed Poisson spike train for each electrode. This eliminates all correlations between spike trains, so that non zero values of $R$ are the result of finite sampling. The Poisson spike train contains the same number of spikes as the experimentally observed spike train. In practice, $R$ values obtained from this model are very small since coincidences are minimal between random Poisson spike trains.  The Poisson model establishes a background noise level below which $R$ values were considered to be effectively 0. This null model was used to establish the background noise level below which $R$ values were considered to be effectively 0.  The mean Poisson $R$ values over 10 instances of the model were on the order $10^{-5}$ while typical values from the experimental data ranged between $-10^{-3}$ and  $10^{-1}$.

\subsection{Results}
\label{results}
We analyze eight developing networks, four dense ($2.5\pm1.5$ cells/mm$^2$) and four sparse ($0.6\pm0.24$ cells/mm$^2$). For each network on each recording day (\ref{table:t1}), new activation graphs are generated by the competitive first response model (see Methods) and directional activation links between pairs are established. Thus, pairs can be represented by the coordinate $[X_n,Y_n]$, where $X_n$ is the estimated probability that electrode $X$ activates $Y$ and $Y_n$ is the estimated probability $Y$ activates $X$ (Fig \ref{fig:f4}b).  
If $X_n$=$Y_n$, the point falls on the line $y=x$ (solid line, Fig. \ref{fig:f4}b). Each coordinate has a magnitude (distance from zero) and a deviation angle from the line $y=x$. If $X_n$ = $Y_n$ , then the deviation angle = $0$ and the magnitude = $\sqrt{X^2+Y^2}$ .
In figures \ref{fig:f4}c \& \ref{fig:f4}d we examine the magnitude and deviation of all non-zero ($X_n \ne 0$ \& $Y_n \ne 0$) coordinates in all networks.
In figure \ref{fig:f4}c, the average coordinate magnitude for dense and sparse networks on each recorded DIV is plotted. 
At early ages, the average coordinate magnitude in dense networks is greater than in their sparse counterparts. 
This indicates that pairwise connections form faster in dense networks. However, in sparse networks, the average coordinate magnitude stabilizes at (slightly) higher values than in dense networks (Fig. \ref{fig:f4}c).  In Fig \ref{fig:f4}d, we examine the average deviation from the line $y=x$ in dense and sparse networks, for all active pairs. Pairwise activation is relatively balanced during  dense network maturation ($X_n\approx Y_n$), as demonstrated by low deviation angles. In sparse networks, deviation angles decrease, but remain larger than dense deviation angles throughout maturation. 

Wagenaar et al. \cite{wagenaar_2006} observed that sparse networks are slower to develop network bursting than their dense counterparts (Fig. \ref{fig:f4}a). Additionally, activation graphs reveal that late onset bursting, a function of pairwise interactions \cite{ham_2008}, is highly correlated to the slow development of pairwise activation connections. Therefore, early development of pairwise interactions in dense networks appears to be related to the onset of multiple network bursts per minute around $10$ DIV. Note that multiple bursts per minute are not a present in sparse networks until about $22$ DIV (Fig \ref{fig:f4}a). 

\subsection{link entropy}
\label{link entropy}
\begin{figure*}[htbp]
    \includegraphics[width=.95\textwidth]{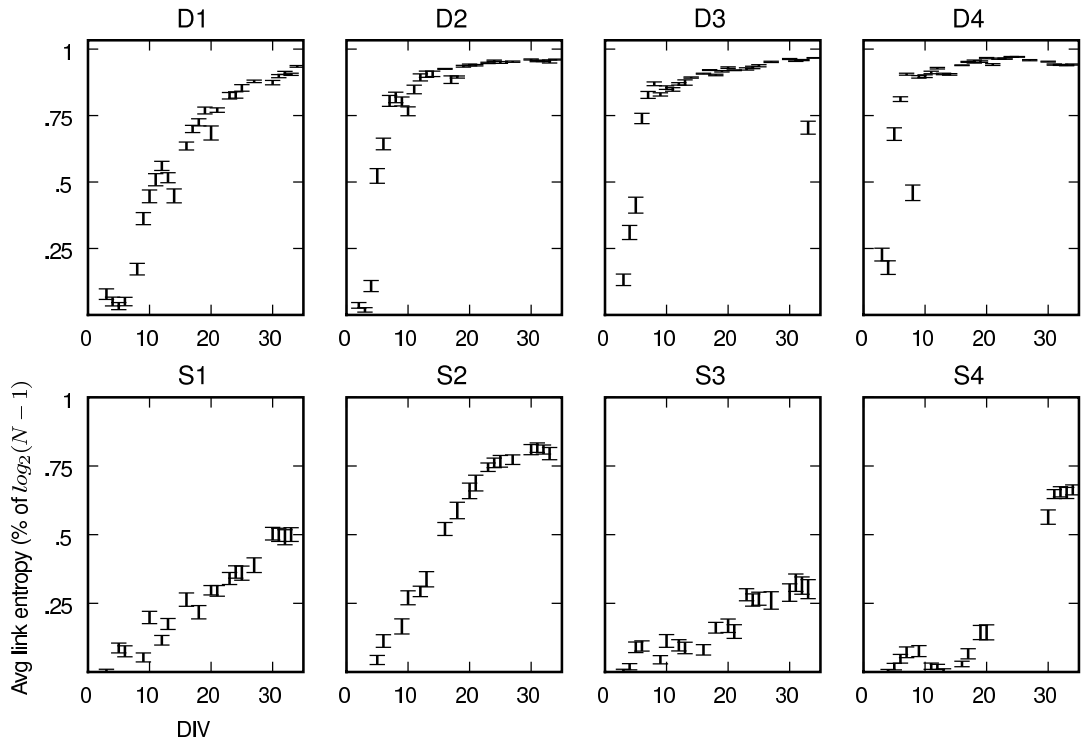}
     \caption{Average daily activation link entropy and standard error for dense ($D1-D4$) and sparse networks ($S1-S4$). 
Averages are plotted as a percent of the maximum link entropy, $log_2(N-1)$.
Average values in dense networks come close to maximal value, indicating that all nodes in the network have nearly equally probable connections with all other nodes.  
Larger, but not maximal link entropy in S2 correlates to the development of redundant groups in Fig. 6.  
The values plotted are normalized by the maximum possible value, which corresponds to $1$ on the $y$ axis \label{fig:f5}}
\end{figure*}

Figure \ref{fig:f5} shows the average daily link entropy for all electrodes in each of the eight networks studied. Note that a similar number of active electrodes were observed in all networks (Table \ref{table:t1}).
As dense networks mature, average individual electrode link entropy values approach the maximum possible value, which indicates nearly uniform probabilistic links between any target electrode and all other active electrodes. In other words, given activation at a particular dense network electrode, the next electrode to activate is almost completely random. Mature sparse networks exhibit midrange link entropy values, indicating that probabilistic connections from a target to all other electrodes are unequally distributed and therefore more predictable. This corresponds to the larger deviation angles observed in sparse networks (Fig. \ref{fig:f4}d). Note that sparse network S2 has larger link-entropy values than S1, S3 or S4 though not as large as in the dense networks.

\subsection{Informational relationships of order 3}
\label{ir3}
Groups of three electrodes (triplets), are the smallest (and most abundant) arrangement capable of providing information about functional connectivity in a network \cite{bettencourt_2008}. As described in Methods, informational structures are determined by comparing the information from two electrodes together conditional on the third with information gained from the two electrodes separately conditioned on the third. The resulting value ($R$) categorizes activity between electrodes as redundant, synergetic or independent. 

\begin{figure*}[htbp]
    \includegraphics[width=.95\textwidth]{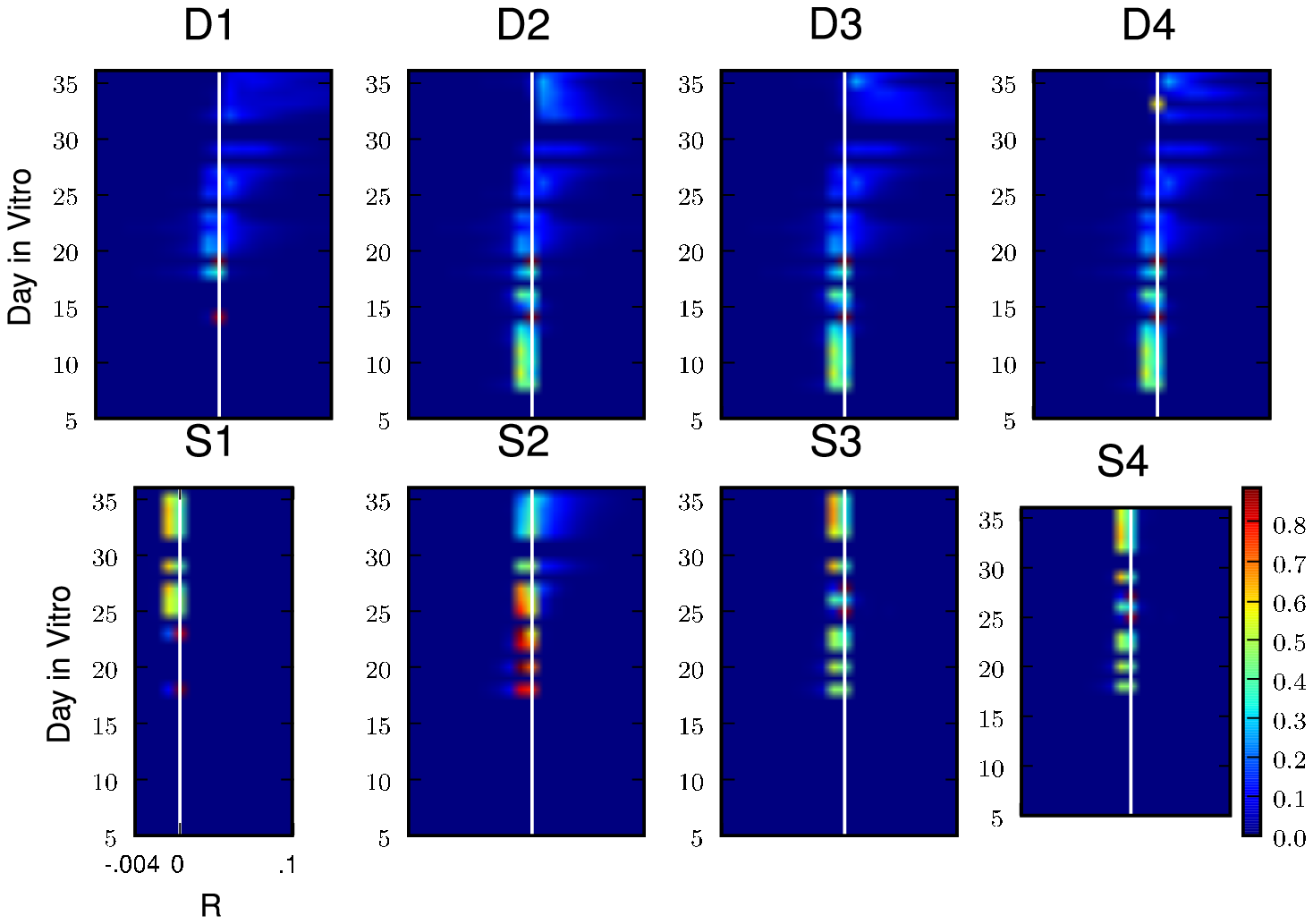}
     \caption{Computed $R$ values for all ensembles of $3$ electrodes in developing networks. Independent $R$ values ($R=0$) are not depicted. Dense networks (D1-D4) are depicted on the top row and sparse (S1-S4) on the bottom row.  
Individual triplets in dense networks change from primarily synergetic relationships to primarily redundant at about $18$ DIV and by $30$ DIV redundant ensembles dominate.  
At early and late stages of maturation, sparse networks S1, S3 and S4 are dominated by synergetic relationships. 
Starting around $18$ DIV, sparse networks switch between primarily redundant and primarily synergetic until about $25$ DIV when they settle on primarily redundant.
Note in S2, redundant ensembles that develop around $28$ DIV though synergetic relationships do not disappear as in dense network \label{fig:f6}}
\end{figure*}

On each recording day, $R$ values for all unique electrode triplet ensembles are calculated (see Methods).
Note that $R(X,Y,Z)=R(X,Z,Y)$. In Figure \ref{fig:f6}, distributions of $R$ values from each network on each recorded day are shown. 
Synergetic values ($R<0$) are grouped in bin sizes of $.001$ bits while redundant values ($R>0$) are grouped in bins of size $.01$ bits. 
The asymmetry in relative sizes of positive and negative $R$ values has been demonstrated previously \cite{bcourt_2007}.  
Independent values ($R=0$) account for 35\% of all $R$ values and are not shown. As described in Methods, all values are obtained during network burst events to provide a comparison with earlier studies.

\begin{figure*}[htbp]
    \includegraphics[width=.50\textwidth]{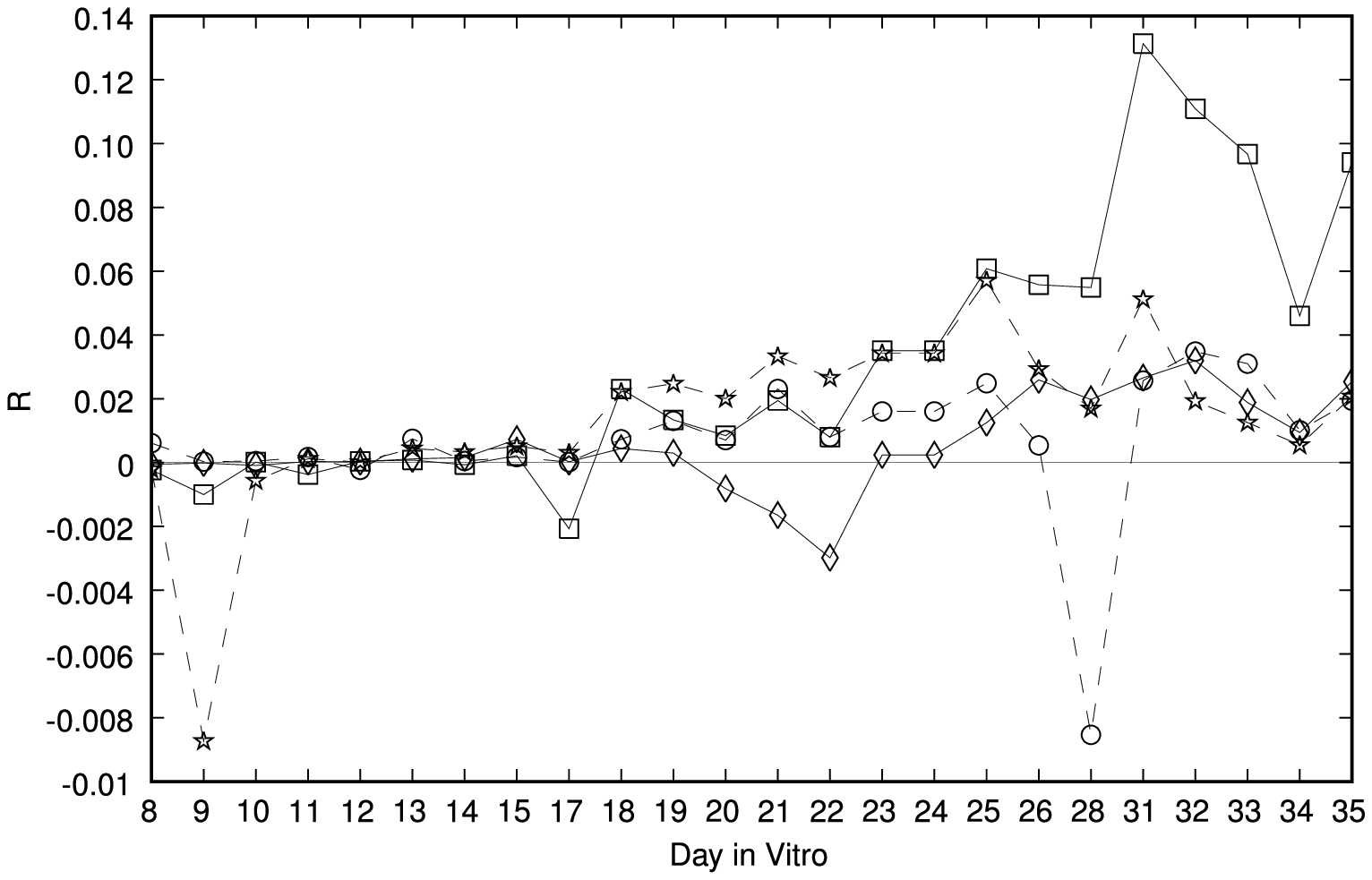}
     \caption{Daily $R$ values of four representative triplets in network $D4$. Note that some triplets switch back and forth between synergetic ($R<0$) and redundant ($R>0$) informational states during development.  
However, after $28$ DIV the vast majority of network triplets in $D4$ were either redundant or independent \label{fig:f7}}
\end{figure*}

In all instances, young network bursts ($10-20$ DIV) are dominated by synergetic triplets. However, around $18$ DIV, all networks show a shift towards redundant triplets. 
By $30$ DIV, matured dense networks are dominated by redundant triplet ensembles. 
This finding is commensurate with our previous study of a mature $45$ DIV network \cite{bcourt_2007}.
Sparse networks are observed to switch back and forth between primarily synergetic and primarily redundant from $18-25$ DIV. Unlike their dense counterparts, sparse network triplets became primarily synergetic after $25$ DIV. 
While three out of the four sparse networks remain primarily synergetic (Fig \ref{fig:f6}, S1, S3, S4), in one network (S2) redundant groups reemerge around $30$ DIV and are present throughout the remainder of the experiment. 
Note that synergetic triplets still maintain a strong presence during this time period. This is in clear contrast to dense networks where synergetic groups essentially disappear altogether in favor of redundant ensembles. 
This shift is primarily due to the evolution of individual triplet $R$ values changing from synergetic to redundant. 
In Figure \ref{fig:f7}, several randomly selected representative triplets and their $R$ values on different recording days are shown from network D4. We observe that  the vast majority of dense network triplets (98\%) that are synergetic at a young age become redundant during maturation.

At around $21$ DIV, redundant ensembles begin emerging in both dense and sparse networks (Fig. \ref{fig:f6}). This correlates to the time period when cortical networks {\it in vitro} are generally considered mature \cite{jimbo_1999} and when Wagenaar et al. \cite{wagenaar_2006} observed that spike rates stopped increasing and leveled off. 

\subsection{Triplet interactions}

Large link-entropy values and the presence of redundant ensembles do not appear to be directly correlated to one another. 
However, large values of link entropy seem to be an indicator that redundant ensembles will dominate the informational relationships. 
It should be noted that in D1, the rise in average link-entropy and the increase in redundant triplets are well correlated, but this is not observed in the other three dense networks where link-entropy maximizes before redundant ensembles fully dominate.
Regardless, results suggest that high values of link entropy are necessary for the emergence and dominance of redundant neuronal ensembles. Figure \ref{fig:f5} S2 indicates that, with moderate link entropy values, synergetic and redundant groups coexist. Relatively low link entropy, like that seen in S1, S3, and S4, appears indicative of synergetic relationships (Fig. \ref{fig:f5}).

\section{Discussion}
\label{discuss}

In this manuscript, we explored quantitative patterns of neuronal interactions in developing dense and sparse cultured cortical networks. Such networks are formed when dissociated prenatal cortical tissue is plated on microelectrode arrays and new connections are spontaneously formed. Multisite electrophysiological data from these networks provides unique access to the development of cortical cultures. It should be noted that {\it in vitro} network formation is guided by similar mechanisms (synaptogenesis; \cite{herndon_1981}) used by intact nervous systems.

\subsection{Pairwise interactions}
\label{pwi}

At the most basic level, network formation can be thought of as a pairwise phenomenon where two neurons form synaptic connections with one another through electrical and chemical means. We extrapolate functional pairwise connections through a competitive first response model which is based on the assumption that all previous activity (within a biologically plausible time window) contributes to action potential initiation. This model is used to create directed, weighted activation graphs.

Mature activation graphs of dense networks reveal that connections between electrodes are mostly reciprocal ($X_n = Y_n$, Fig \ref{fig:f4}d) as indicated by relatively low deviation angles. In contrast, electrodes in sparse networks tend to develop activation connection strengths that are skewed in one direction or the other (larger deviation angle, $X_n \ne Y_n$, Fig. \ref{fig:f4}d). Additionally, link entropy in sparse networks is smaller than link entropy in dense networks (S1-S4, Fig. \ref{fig:f5}). Together, these results indicate that electrodes show activation pathway biases (low link entropy) in sparse networks, but these biases are not reciprocal between pairs.  In dense networks, the opposite appears true; probabilistic activation pathways between electrodes are nearly uniform (large values of link entropy, Fig. \ref{fig:f5}) and reciprocal between pairs. 

In networks processing information, each type of connection structure has advantages and disadvantages. 
Dense networks appear to have more redundancy (high link entropy) which may allow for higher fault tolerance: if a path via a given electrode becomes unresponsive to activation attempts by another, many other pathways could easily be activated.  
However, such a system may be detrimental to rapid information processing where fewer strong pathways, like those seen in sparse networks, could help an animal reach a fast decision state with greater precision. 
The presence of favored activation pathways in sparse networks may indicate that they are better platforms for performing training studies that seek to change network interactions {\it in vitro} \cite{shahaf_2001}. 
Since there are fewer probabilistic pathways for modification, changes to these connections will likely produce greater differences in existing efferent and afferent activation pathways. 
Conversely, shutting down or building pathways in dense networks would likely produce a less noticeable change in directional activation patterns as many detours are available.

Ensembles of $3$ nodes are the minimum required to form computational groups \cite{bettencourt_2007}. Sequential chains of redundant cells, shown in Figure \ref{fig:f3}b, can be used to relay information, and similar arrangements have been shown to play a role in short-term memory in the brain \cite{abeles_1991} \cite{diesmann_1999}. 
In the synergetic configuration depicted in Fig \ref{fig:f2}a, a neuron receives inputs from many other cells in such a way that spiking activity is a nonlinear function of the inputs. 
Both systems (redundant and synergetic) allow neurons to integrate activity from many sources and process information. 

We demonstrated that plating density has an important effect on informational relationships. 
Dense and sparse networks develop similar distributions of synergetic informational structures between $10$ and $20$ DIV (Fig. \ref{fig:f6}). However, as dense networks mature, most triplets change from synergetic to redundant functional informational connections (Fig. \ref{fig:f7}). 
By about $30$ DIV very few synergetic triplets remain in dense networks. 
Conversely, individual triplets in sparse networks remain primarily synergetic from early to late stages of development. 
These results suggest that sparse networks may be better suited for processing information since synergetic relationships are necessary to perform these tasks \cite{bcourt_2007}. 
Of course synergetic relationships also exist in denser networks, but in a smaller relative proportion. They may also occur in larger groups of cells, not analyzed here \cite{bettencourt_2008}.
One mature sparse network, S2 (Fig. \ref{fig:f6}), showed a relatively even mix of both synergetic and redundant ensembles during late stages of maturation. This was not seen in any of the other networks. 
Furthermore, this network exhibited link entropy values lying between the other sparse networks and the dense networks (Fig \ref{fig:f5}).  
These findings suggest a link between activation connections and informational relationships.  
Specifically, when activation connections are balanced throughout a network (larger link entropy values), redundant ensembles result. 
However, if activation connection are not quite balanced (mid range link entropy values), but are close, both synergetic and redundant ensembles appear.

The observed density dependent functional organization of cortical tissue {\it in vitro} raises interesting questions for future research. To elucidate how network density affects connectivity structures, experiments that span a broader range of plating densities are needed.  Additionally, it would be interesting to apply these analytical techniques to recordings where a network is monitored continuously during development.  

\subsection{Conclusion}
\label{conc}

The development of functional connectivity in neural networks {\it in vitro} comprising similar cells is greatly affected by plating density. 
In a comparison of dense and sparsely prepared networks, we demonstrate that strong pairwise connections occur earlier in dense networks.
Link entropy analysis of pairwise activation graphs reveal that nodal connections tend to be biased in sparse and nearly equal in dense networks.
Connections between electrode triplets in both dense and sparse networks are primarily synergetic during early stages of development. 
However, in dense networks, they become primarily redundant by $30$ DIV. Conversely, in sparse networks triplets are primarily synergetic by $30$ DIV. 
Due to the fact that {\it in vitro} cultures have many similarities with {\it in vivo} tissue, we believe that the developmental features identified here may also play a role in living organisms. 
These findings should be applicable to future research seeking to replicate or emulate brain functions and provide useful constraints for experiments of neural function.
\section{acknowledgements}

 We are grateful to Daniel Wagenaar, Jerome Pine and Steve Potter for making the data analyzed in this manuscript available to us. This work was supported by the Los Alamos National Laboratory Directed Research and Development (LDRD) 2009006DR synthetic cognition through petascale models of the primate visual cortex project.

\bibliographystyle{plain}
\bibliography{JCNSfile}


\end{document}